# Exact solutions to the nonlinear Schrödinger equation with time-dependent coefficients

Xin-Lei Mai, and Wei Li

School of Mathematics and Physics, Beijing University of Chemical Technology, Beijing 100029, China

Correspondence should be addressed to Xin-Lei Mai: maixinlei7@163.com

Correspondence should be addressed to Wei Li: liwei2@mail.buct.edu.cn

## Abstract

In this paper, the trial function method is employed to find the exact solutions for high-order nonlinear Schrödinger equations with time-dependent coefficients. This system describes the propagation of ultrashort light pulses in nonlinear fibers, with self-steepening and self-frequency shift effects. As a result, we derive a range of exact solutions which include Jacobi elliptic function solutions, solitary wave solutions, and rational function solutions.

## I. Introduction

Many physical phenomena in nature can be described by the nonlinear Schrödinger equation (NLSE), which widely exists in many fields such as plasma physics [1], fluid dynamics [2], nonlinear optics [3], quantum mechanics [4], hydrodynamics [5], biology, and many more. Especially, NLSE is a general model to describe the propagation of light pulses in optical fibers. Therefore, the study of solitary wave solutions of NLSE has important theoretical and practical significance for understanding the nature of physical phenomena described by NLSE.

In recent decades, many effective methods have been proposed to investigate the exact solutions, such as the homogeneous balance method [6], the tanh function expansion method and its extension [7], the sine-cosine methods [7], the exp-function method [7], the multiple exp-function method [8], the first integral method [9], the Jacobi elliptic function expansion method [10], the sub-ODE method [11], the $(G'/G)$- expansion method [12], the modified simple equation method [13], the extended auxiliary equation method [14], the $\exp(-\phi(\xi))$- expansion method [15], and the trial function method and its extension [16].

For all we know, most of the aforementioned methods are related to constant coefficients models. However, solving variable coefficient NLSE is much more difficult than their constant coefficient counterparts because of the existence of their coefficients are time-dependent functions.

Recently, much attention has been paid to variable coefficient NLSE, and many approaches have been extended to solve these models [16-22]. Liu [21] proposed a trial function method which can be suitable to both real equations and complex equations with variable coefficients.

In this paper, the trial function method is applied to find the exact solutions of the following cubic-quintic NLS equation with time-dependent coefficients

$$iq_t + f(t)q_{xx} + g(t)\left(|q|^2 + \sigma|q|^4\right)q = ih(t)\left(|q|^2 q\right)_x + ip(t)\left(|q|^2\right)_x q \qquad (1)$$

where $q(x,t)$ is the complex envelope of the electric field, $x$ and $t$ are the distance along the direction of propagation and time, respectively. $f(t)$ is the dispersion coefficient, $h(t)$ is the self-steepening coefficient, $p(t)$ is the self-frequency shift coefficient, and $\sigma$ is a constant. This variable coefficient Schrödinger equation with high-order dispersion terms and high-order nonlinear terms has attracted a great deal of attention. It describes the pulse transmission in the femtosecond state and considers the loss in the transmission process [17,22]. Green and Biswas [22] studied Eq. (1) by the ansatz method and obtained the exact soliton solution under certain parametric conditions. We use the trial function method to find its exact solutions.

This paper is organized as follows. In Section 2, the trial function method has been employed to simplify Eq. (1). In Section 3, we use the suitable transformation for constructing exact solutions of Eq. (1) and numerically simulate the solutions by selecting appropriate parameters. Section 4 contains the conclusions of our results.

## II. Trial function method

Assume that the solution of Eq. (1) can be expressed as

$$q(x,t) = u(\xi)e^{i\eta}, \xi = k(t)x + w(t), \eta = s(t)x + r(t) \qquad (2)$$

where $u$ is a function of $\xi$ to be determined later, and $k(t)$, $w(t)$, $s(t)$, and $r(t)$ are undetermined parameters related to time. Substituting Eq. (2) into Eq. (1) and separating the real and the imaginary parts, we have

$$[k'(t)x + w'(t) + 2f(t)s(t)k(t)]u' - [3h(t) + 2p(t)]k(t)u^2 u' = 0 \qquad (3)$$

$$f(t)k^2(t)u'' - [s'(t)x + r'(t) + f(t)s^2(t)]u + [g(t) + h(t)s(t)]u^3 + \sigma g(t)u^5 = 0 \qquad (4)$$

Considering the structure of the equation, we suppose that the solution satisfies

$$(u')^2 = F(\xi) = \sum_{i=0}^{m} a_i u^i \qquad (5)$$

where $a_i (i = 0, \cdots, m)$ are constants and $m$ is an integer to be determined later. The value of $m$ in Eq. (5) is determined as $m = 6$ by using the homogeneous balance theory. Substituting Eq. (5) into Eq. (4) and setting each coefficient of $u'$, $u^2 u'$ in Eq. (3) and $u^i$ in Eq. (4) to zero, we obtain the system of algebraic equations in the following form

$$\begin{cases} k'(t)x + w'(t) + 2f(t)s(t)k(t) = 0, \quad [3h(t) + 2p(t)]k(t) = 0, \quad a_1 = a_3 = a_5 = 0, \\ a_2 = \dfrac{s'(t)x + r'(t) + f(t)s^2(t)}{f(t)k^2(t)}, \quad a_4 = -\dfrac{g(t) + h(t)s(t)}{2f(t)k^2(t)}, \quad a_6 = -\dfrac{\sigma g(t)}{3f(t)k^2(t)}. \end{cases} \quad (6)$$

Setting

$$k(t) = k \neq 0, \ s(t) = s \neq 0, \ r(t) = c_1 \int f(t)dt, \ g(t) = c_2 k^2 f(t), \ h(t) = c_3 k^2 f(t), \ a_0 = c_4$$

where $k$, $s$, $c_1$, $c_2$, $c_3$, and $c_4$ are arbitrary constants, $f(t)$ is an arbitrary function. The other coefficients can be determined in the following form

$$p(t) = -\frac{3}{2} c_3 k^2 f(t), \ w(t) = -2sk \int f(t) dt,$$

$$a_1 = a_3 = a_5 = 0, \ a_2 = \frac{c_1 + s^2}{k^2}, \ a_4 = -\frac{c_2 + sc_3}{2}, \ a_6 = -\frac{\sigma c_2}{3}. \quad (7)$$

Under the conditions (7), Eq. (5) with m = 6 can be reduced to the following expression

$$(u')^2 = a_0 + a_2 u^2 + a_4 u^4 + a_6 u^6. \quad (8)$$

## III. Exact solutions to equation (1)

The case of $a_0 = 0$ is discussed in [17] and the structure of four soliton solutions appears in the system (1) as follows:

(1) $a_4^2 - 4a_2 a_6 > 0, a_2 > 0$, Eq. (1) has a bright soliton solution:

$$q = \left\{ \frac{2a_2}{\varepsilon \sqrt{a_4^2 - 4a_2 a_6} \cosh(2\sqrt{a_2} \xi) - a_4} \right\}^{\frac{1}{2}} \cdot e^{i\eta}.$$

(2) $a_4^2 - 4a_2 a_6 < 0, a_2 > 0$, Eq. (1) has a singular soliton solution:

$$q = \left\{ \frac{2a_2}{\varepsilon \sqrt{4a_2 a_6 - a_4^2} \sinh(2\sqrt{a_2} \xi) - a_4} \right\}^{\frac{1}{2}} \cdot e^{i\eta}.$$

(3) $a_4^2 - 4a_2 a_6 = 0, a_2 > 0$, Eq. (1) has a dark soliton solution:

$$q = \left\{ -\frac{a_2}{a_4} \left[ 1 + \varepsilon \tanh\left( \frac{\sqrt{a_2}}{2} \xi \right) \right] \right\}^{\frac{1}{2}} \cdot e^{i\eta}.$$

(4) $a_2 > 0$, Eq. (1) has another bright soliton-like solution:

$$q = \left\{ \frac{-a_2 a_4 \sec h^2\left(\sqrt{a_2}\xi\right)}{a_4^2 - a_2 a_6\left[1 + \varepsilon \tanh\left(\sqrt{a_2}\xi\right)\right]^2} \right\}^{\frac{1}{2}} \cdot e^{i\eta}.$$

where $\varepsilon = \pm 1$.

This paper continues to discuss the structure of the soliton solution of the system (1) in the case where $a_0 \neq 0$. In order to exploit the general solution, we use the transformation $\varphi = u^{-2}$ to Eq. (8). Thus, Eq. (8) can be transformed to

$$\varphi' = \pm 2\sqrt{a_0\left(\varphi^3 + \frac{a_2}{a_0}\varphi^2 + \frac{a_4}{a_0}\varphi + \frac{a_6}{a_0}\right)} \qquad (9)$$

With respect to the solution of Eq. (8), we note

$$G(\varphi) = \varphi^3 + \frac{a_2}{a_0}\varphi^2 + \frac{a_4}{a_0}\varphi + \frac{a_6}{a_0}$$

and

$$A = \frac{a_2^2 - 3a_0 a_4}{a_0^2}, \quad B = \frac{a_2 a_4 - 9a_0 a_6}{a_0^2}, \quad C = \frac{a_4^2 - 3a_2 a_6}{a_0^2}, \quad \Delta = B^2 - 4AC.$$

The Eq.(9) can be divided into the following three cases depending on the sign of $\Delta$ by Shengjin discrimination method:

Case 1: $\Delta < 0$, $G(\varphi)$ has three zeros $\varphi_1 < \varphi_2 < \varphi_3$, $\varphi_1 = -\frac{a_2}{3a_0} - \frac{2}{3}\sqrt{A}\cos\frac{\theta}{3}$,

$\varphi_{2,3} = -\frac{a_2}{3a_0} + \frac{\sqrt{A}}{3}\left(\cos\frac{\theta}{3} \pm \sqrt{3}\sin\frac{\theta}{3}\right)$, where $\theta = \arccos T$ and $T = \frac{2Aa_2 - 3a_0 B}{2a_0\sqrt{A^3}}$.

Then Eq. (9) can be converted to

$$\varphi' = \pm 2\sqrt{a_0(\varphi - \varphi_1)(\varphi - \varphi_2)(\varphi - \varphi_3)}$$

Thus, the Eq. (1) has a periodic wave solution given by

$$q = \begin{cases} \dfrac{1}{\left[\varphi_1 cn^2\left(\varepsilon\sqrt{a_0(\varphi_3 - \varphi_1)}\xi, m_1\right) + \varphi_2 sn^2\left(\varepsilon\sqrt{a_0(\varphi_3 - \varphi_1)}\xi, m_1\right)\right]^{\frac{1}{2}}} \cdot e^{i\eta} & a_0 > 0 \\[2ex] \dfrac{sn\left(\varepsilon\sqrt{-a_0(\varphi_3 - \varphi_1)}\xi, m_2\right)}{\left[\varphi_1 - \varphi_3 cn^2\left(\varepsilon\sqrt{-a_0(\varphi_3 - \varphi_1)}\xi, m_2\right)\right]^{\frac{1}{2}}} \cdot e^{i\eta} & a_0 < 0 \end{cases}$$

where $m_1 = \sqrt{\dfrac{\varphi_2 - \varphi_1}{\varphi_3 - \varphi_1}}$ and $m_2 = \sqrt{\dfrac{\varphi_3 - \varphi_2}{\varphi_3 - \varphi_1}}$.

We select $f(t) = 1$, $f(t) = t$, and $f(t) = 1/t$, respectively. The corresponding periodic waveform is showed in Figure 1.

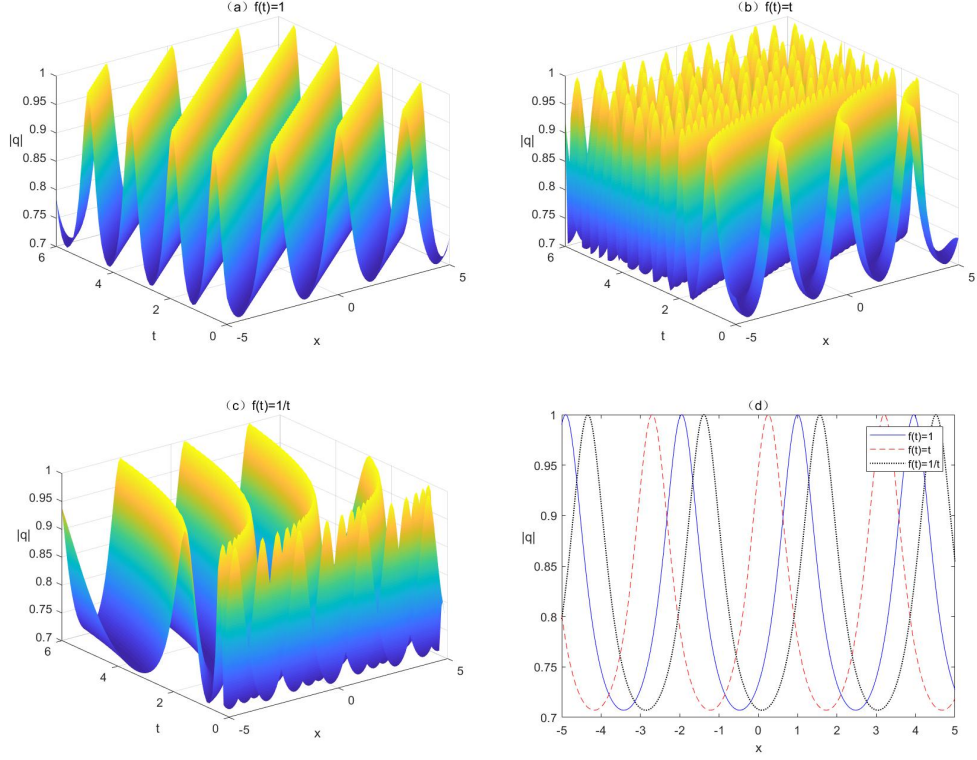

Figure 1: (a)-(c) The 3D wave profile of |q|. (d) The wave along the x axis.

Case 2: $\Delta > 0$, $G(\varphi)$ has one zero $\varphi_1$, $\varphi_1 = -\dfrac{a_2}{3a_0} - \dfrac{1}{3}\left(\sqrt[3]{Y_1} + \sqrt[3]{Y_2}\right)$, where

$$Y_{1,2} = A\dfrac{a_2}{a_0} + 3\left(\dfrac{-B \pm \sqrt{B^2 - 4AC}}{2}\right).$$

Then Eq. (9) can be converted to

$$\varphi' = \pm 2\sqrt{a_0(\varphi - \varphi_1)(\varphi^2 + \alpha\varphi + \beta)}$$

where $\alpha = \dfrac{2a_2}{3a_0} - \dfrac{1}{3}\left(\sqrt[3]{Y_1} + \sqrt[3]{Y_2}\right)$ and $\beta = \dfrac{1}{36}\left(\sqrt[3]{Y_1} + \sqrt[3]{Y_2} - \dfrac{a_2}{3a_0}\right)^2 + \dfrac{1}{12}\left(\sqrt[3]{Y_1} - \sqrt[3]{Y_2}\right)^2$.

A periodic solution with an elliptic function in Eq. (1) is given by

$$q = \begin{cases} \left[\dfrac{1+cn\left(2\varepsilon\sqrt{a_0}\left(\varphi_1^2+\alpha\varphi_1+\beta\right)^{\frac{1}{4}}\xi, m_3\right)}{\left(\varphi_1+\sqrt{\varphi_1^2+\alpha\varphi_1+\beta}\right)+\left(\varphi_1-\sqrt{\varphi_1^2+\alpha\varphi_1+\beta}\right)cn\left(2\varepsilon\sqrt{a_0}\left(\varphi_1^2+\alpha\varphi_1+\beta\right)^{\frac{1}{4}}\xi, m_3\right)}\right]^{\frac{1}{2}} \cdot e^{i\eta} & a_0 > 0 \\[2em] \left[\dfrac{1+cn\left(2\varepsilon\sqrt{-a_0}\left(\varphi_1^2+\alpha\varphi_1+\beta\right)^{\frac{1}{4}}\xi, m_4\right)}{\left(\varphi_1-\sqrt{\varphi_1^2+\alpha\varphi_1+\beta}\right)+\left(\varphi_1+\sqrt{\varphi_1^2+\alpha\varphi_1+\beta}\right)cn\left(2\varepsilon\sqrt{-a_0}\left(\varphi_1^2+\alpha\varphi_1+\beta\right)^{\frac{1}{4}}\xi, m_4\right)}\right]^{\frac{1}{2}} \cdot e^{i\eta} & a_0 < 0 \end{cases}$$

where $m_3 = \dfrac{1}{2}\left(2 - \dfrac{2\varphi_1+\alpha}{\sqrt{\varphi_1^2+\alpha\varphi_1+\beta}}\right)^{\frac{1}{2}}$ and $m_4 = \dfrac{1}{2}\left(2 + \dfrac{2\varphi_1+\alpha}{\sqrt{\varphi_1^2+\alpha\varphi_1+\beta}}\right)^{\frac{1}{2}}$.

The corresponding waveform of periodic wave is simulated in Figure 2 by selecting three functional types of $f(t)=1$, $f(t)=t$ and $f(t)=1/t$. This waveform is slightly different from that of Figure 1 and is sharp at the trough.

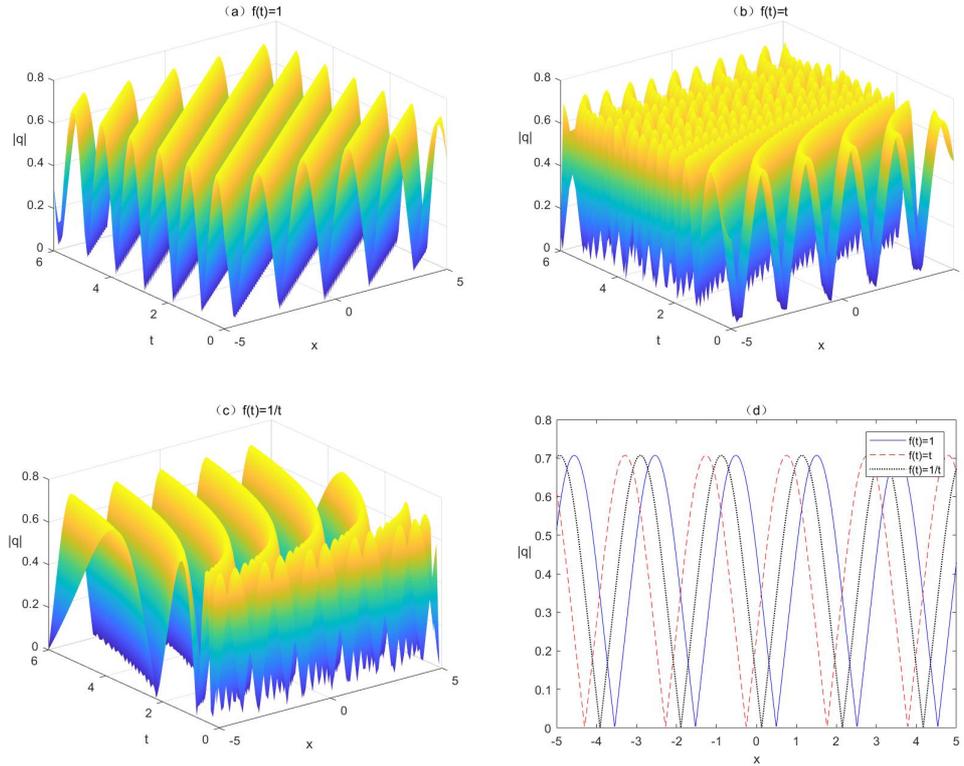

Figure 2: (a)-(c) The 3D wave profile of |q|. (d) The wave along the x axis.

Case 3: $\Delta = 0$

3.1: $A \neq 0$, $G(\varphi)$ has two zeros, in which one is double zero, $\varphi_1 = -\dfrac{K}{2}, \varphi_2 = -\dfrac{a_2}{a_0} + K$, where

$$K = \frac{B}{A}.$$

Then Eq.(9) can be converted to

$$\varphi' = \pm 2\sqrt{a_0(\varphi - \varphi_1)^2(\varphi - \varphi_2)}$$

Thus, the exact solution to Eq.(1) can be written as

$$q = \begin{cases} \dfrac{1}{\left[\varphi_2 \sec^2\left(\sqrt{a_0(\varphi_2 - \varphi_1)}\xi\right) - \varphi_1 \tan^2\left(\sqrt{a_0(\varphi_2 - \varphi_1)}\xi\right)\right]^{\frac{1}{2}}} \cdot e^{i\eta} & a_0\varphi_1 < a_0\varphi_2 \\ \dfrac{1}{\left[\varphi_2 \text{sech}^2\left(\sqrt{a_0(\varphi_1 - \varphi_2)}\xi\right) + \varphi_1 \tanh^2\left(\sqrt{a_0(\varphi_1 - \varphi_2)}\xi\right)\right]^{\frac{1}{2}}} \cdot e^{i\eta} & a_0\varphi_1 > a_0\varphi_2 \end{cases}$$

We select $f(t)=1$, $f(t)=t$, and $f(t)=1/t$ three function types for numerical simulation. For $a_0\varphi_2 > a_0\varphi_1$, the system appears a periodic waveform which is similar to Figure 2 (see Figure 3). For $a_0\varphi_2 < a_0\varphi_1$, the system presents a bright soliton waveform at $a_0 > 0$ (see Figure 4); and a dark soliton waveform at $a_0 < 0$ (see Figure 5). This is the contrast phenomenon that has not appeared in other cases due to the different signs of $a_0$.

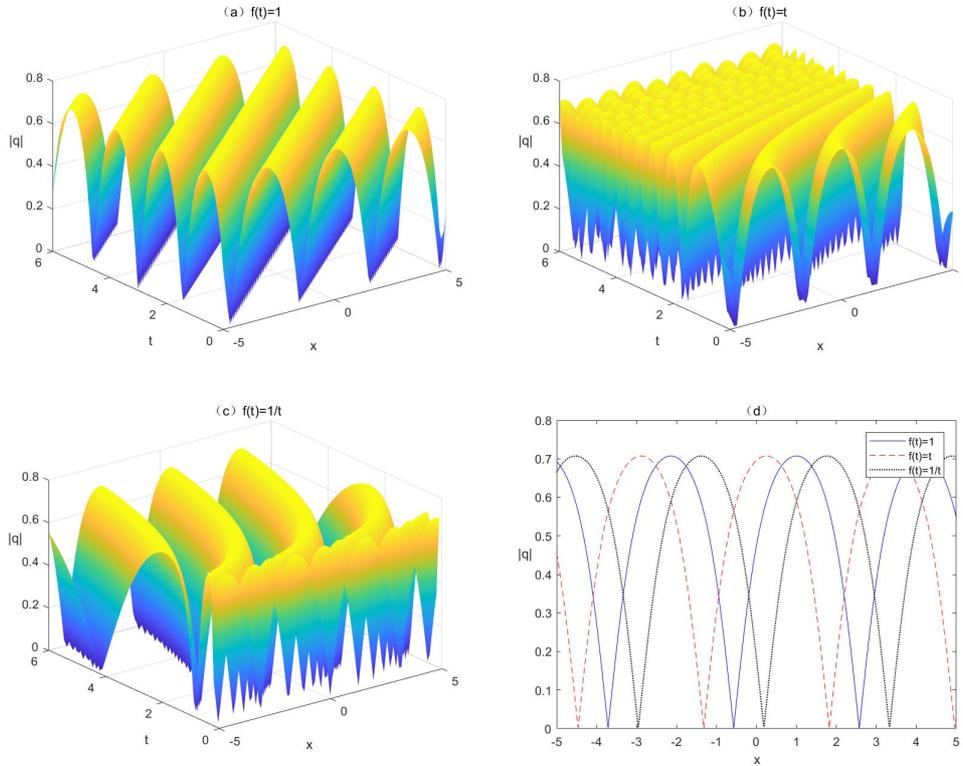

Figure 3: (a)-(c) The 3D wave profile of |q|. (d) The wave along the x axis.

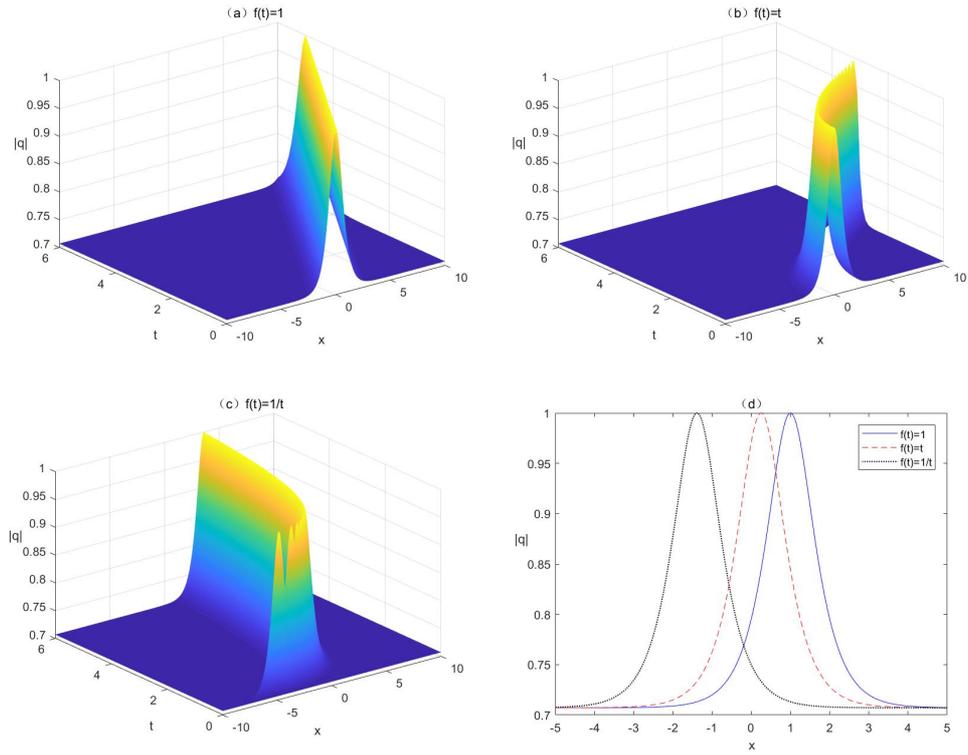

Figure 4: (a)-(c) The 3D wave profile of |q|. (d) The wave along the x axis.

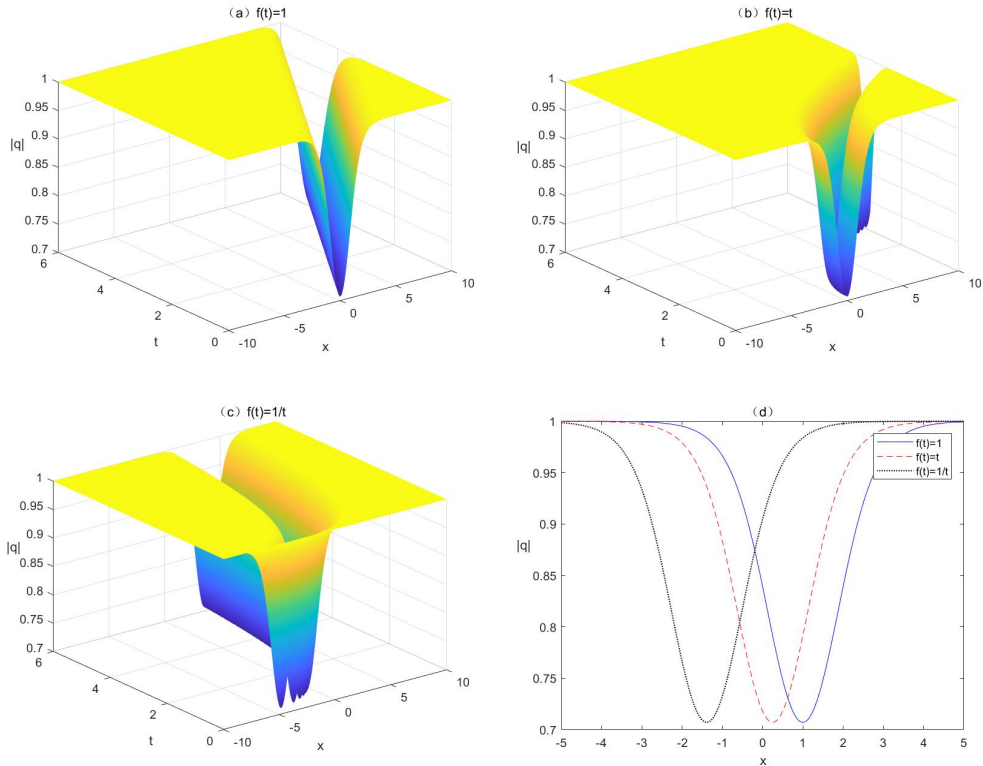

Figure 5: (a)-(c) The 3D wave profile of |q|. (d) The wave along the x axis.

3.2: $A=0$, then $B=0$ i.e. $a_4 = \dfrac{a_2^2}{3a_0}, a_6 = \dfrac{a_2^3}{27a_0^2}$. At this time $G(\varphi)$ has triple zero $\varphi_1 = -\dfrac{a_2}{3a_0}$.

Then Eq.(9) can be converted to

$$\varphi' = \pm 2\sqrt{a_0(\varphi-\varphi_1)^3}$$

The Eq.(1) has a rational function solution given by

$$q = \left(\frac{3a_0\xi^2}{3-a_2\xi^2}\right)^{\frac{1}{2}} \cdot e^{i\eta}$$

For three form of $f(t)$, $f(t)=1$, $f(t)=t$, $f(t)=1/t$, the systems have singular solitary waveform (see Figure 6).

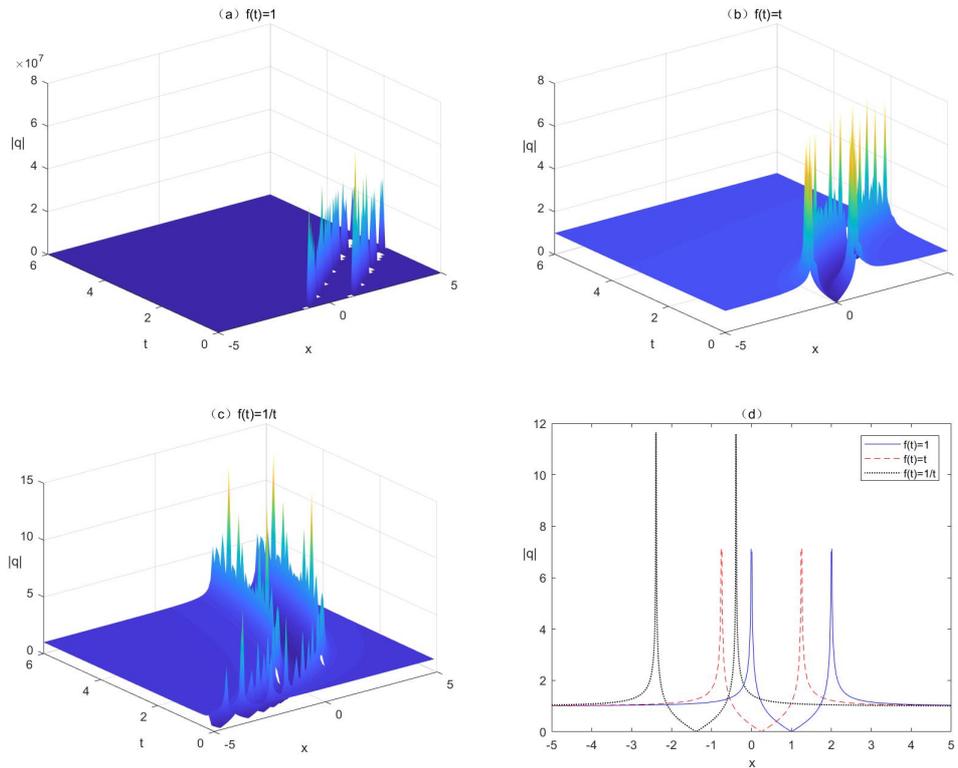

Figure 6: (a)-(c) The 3D wave profile of |q|. (d) The wave along the x axis.

## IV. Conclusions

In this paper, we have investigated a class of high-order nonlinear Schrodinger equation with time-dependent coefficients, simulating the propagation of ultrashort light pulses in nonlinear optical fibers. The trial function method is used to analyze the nonlinear equation with variable coefficients and to find its exact solution. To better describe the dynamic behavior and properties of these solutions, we select three different function types $f(t)=1$, $f(t)=t$, $f(t)=1/t$ to simulate the waveform structure. The results of the analysis of nonlinear equations in this paper reflect the diversity of solitary wave solution structures and have certain research significance. The trial function method is also an effective systematic method for solving nonlinear equations by the linear method.